\documentclass[twocolumn,final,times,sort&compress]{elsarticle}
\usepackage{graphicx,amssymb,amsmath,color}
\usepackage{hyperref}
\journal{Physica B}

\begin{document}

\begin{frontmatter}

\title{Frustrated honeycomb-lattice bilayer quantum antiferromagnet\\ 
       in a magnetic field\tnoteref{grant}}
\tnotetext[grant]{O.~D. and J.~R. acknowledge the support by the Deutsche Forschungsgemeinschaft (project RI615/21-2). 
                  O.~D. acknowledges the kind hospitality of the University of Magdeburg in April-May of 2017.
                  T.~K. and O.~D. were partially supported by Project FF-30F (No. 0116U001539) from the Ministry of Education and Science of Ukraine.}
\author[ICMP,IFNU]{Taras Krokhmalskii}
\author[ICMP]{Vasyl Baliha\corref{coraut}} 
\cortext[coraut]{Corresponding author}
\ead{baliha@icmp.lviv.ua}
\author[ICMP,IFNU,OVGU1]{Oleg Derzhko}
\author[OVGU2]{J\"{o}rg Schulenburg}
\author[OVGU1]{Johannes Richter}
\address[ICMP]{Institute for Condensed Matter Physics, National Academy of Sciences of Ukraine, Svientsitskii Street 1, 79011 L'viv, Ukraine}
\address[IFNU]{Department for Theoretical Physics, Ivan Franko National University of L'viv, Drahomanov Street 12, 79005 L'viv, Ukraine}
\address[OVGU1]{Institut f\"{u}r theoretische Physik, Otto-von-Guericke Universit\"{a}t Magdeburg, P.O. Box 4120, 39016 Magdeburg, Germany}
\address[OVGU2]{Universit\"{a}tsrechenzentrum, Otto-von-Guericke Universit\"{a}t Magdeburg, P.O. Box 4120, 39016 Magdeburg, Germany}

\begin{abstract}
Frustrated bilayer quantum magnets have attracted attention as flat-band spin systems with unconventional thermodynamic properties.   
We study the low-temperature properties 
of a frustrated honeycomb-lattice bilayer spin-$\frac{1}{2}$ isotropic ($XXX$) Heisenberg antiferromagnet in a magnetic field
by means of an effective low-energy theory using exact diagonalizations and quantum Monte Carlo simulations.
Our main focus is on the magnetization curve and the temperature dependence of the specific heat
indicating a finite-temperature phase transition in high magnetic fields.  
\end{abstract}


\end{frontmatter}

\section{Introduction}

\begin{figure}
\begin{center}
\includegraphics[width=0.45\textwidth]{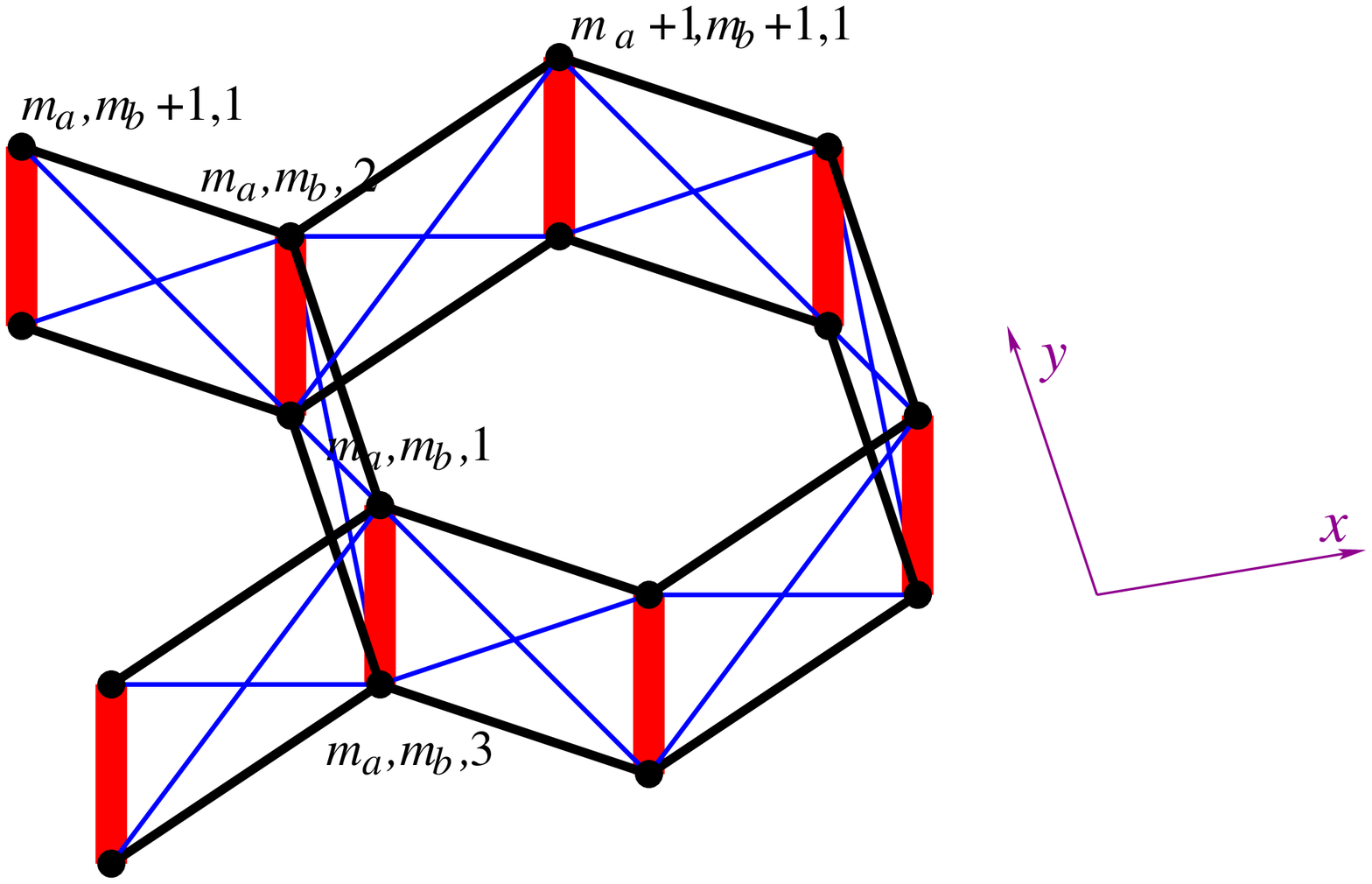}\\
\vspace{5mm}
\includegraphics[width=0.35\textwidth]{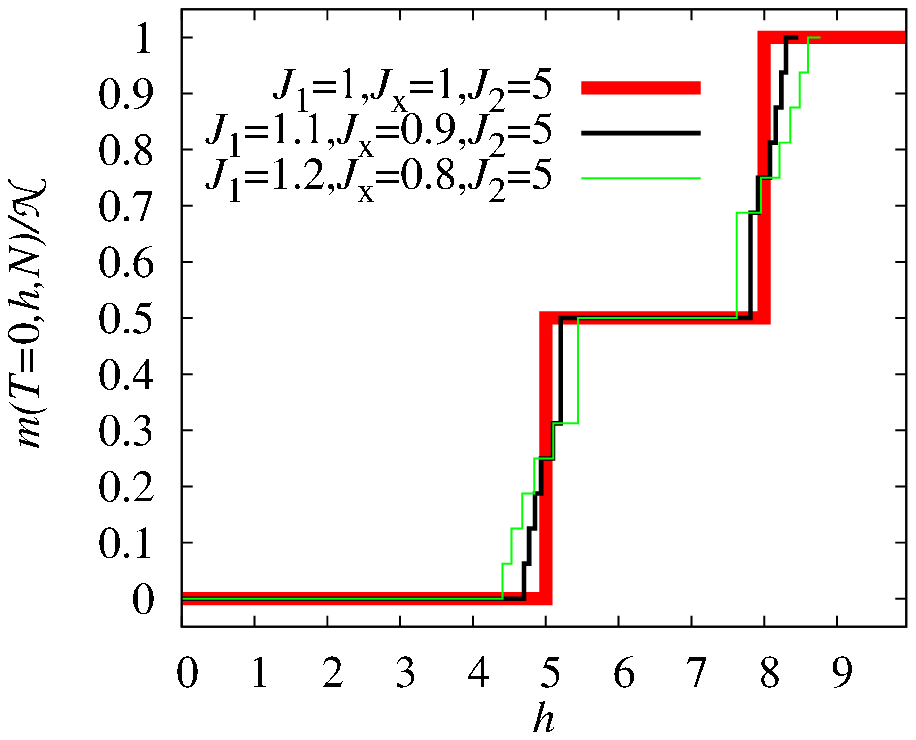}
\end{center}
\vspace{-0.5cm}
\caption{Top: the frustrated honeycomb-lattice bilayer studied in the present paper. 
The unit cell contains 4 sites and the corresponding Bravais lattice is the triangular lattice 
with basis vectors
${\bf{a}}=\sqrt{3}a_0{\bf{i}}$
and
${\bf{b}}=-\frac{\sqrt{3}}{2}a_0{\bf{i}}+\frac{3}{2}a_0{\bf{j}}$
($a_0$ is the hexagon side length).
The integer numbers $m_a$ and $m_b$ determine the position of the unit cell.
The thick red vertical bonds represent the strongest interlayer coupling $J_2$, 
the thin black bonds within each layer correspond to the nearest-neighbor intralayer coupling $J_1$,
and 
the thin blue bonds between the layers correspond to the frustrated interlayer coupling $J_{\rm{X}}$.
Bottom: exact-diagonalization data of the ground-state magnetization curve 
of a finite honeycomb-lattice bilayer of $N=32$ sites for model (\ref{01}) with $J_2=5$
and
$J_1=J_{\rm{X}}=1$ (thick red line),
$J_1=1.1$, $J_{\rm{X}}=0.9$ (thin black line),
$J_1=1.2$, $J_{\rm{X}}=0.8$ (very thin green line).}
\label{fig1}
\end{figure}

In the present paper,
we consider a spin-$\frac{1}{2}$ antiferromagnetic Heisenberg model on a $N$-site two-dimensional lattice shown in Fig.~\ref{fig1}.
The Hamiltonian of the model reads
\begin{eqnarray}
\label{01}
H=\sum_{\langle ij\rangle} J_{ij} {\bf{s}}_i \cdot {\bf{s}}_j -hS^z,
\;
J_{ij}>0,
\;
S^z=\sum_i s_i^z.
\end{eqnarray}
The first sum in Eq.~(\ref{01}) runs over all bonds of the frustrated honeycomb-lattice bilayer,
see Fig.~\ref{fig1},
that is,
$J_{ij}$ acquires three values: 
$J_2$ (vertical red bonds),
$J_1$ (nearest-neighbor intralayer black bonds),
and
$J_{\rm{X}}$ frustrating interlayer blue bonds).
We are interested in the regime 
when $J_2$ is the strongest bond and a deviation of $J_{\rm{X}}$ from $J_1$ is small,
or, more precisely,
$J_2>3J$ with $J=(J_1+J_{\rm{X}})/2$ and $\vert J_1-J_{\rm{X}}\vert/J_2\ll 1$.
If $J_1=J_{\rm{X}}$ one faces the so-called ideal frustration case
characterized by a flat-one magnon band \cite{schulenburg2002,zhitomirsky2004,derzhko2015},
otherwise the system is slightly away from the ideal frustration region in the parameter space.

The described model has attracted some interest recently from experimental and theoretical sides.
On one hand,
the interest in this model stems from experiments on Bi$_3$Mn$_4$O$_{12}$(NO$_3$),
in which the ions Mn$^{4+}$ form a frustrated spin-$\frac{3}{2}$ bilayer honeycomb lattice \cite{smirnova2009,okubo2010,matsuda2010,kandpal2011,alaei2017}.
On the other hand,
there are a few theoretical papers considering the ground-state and low-temperature properties of a Heisenberg antiferromagnet on a frustrated bilayer honeycomb lattice \cite{albarracin2016,zhang2016,krokhmalskii2017},
in which 
classical spin \cite{albarracin2016} or quantum spin-$\frac{1}{2}$ \cite{zhang2016,krokhmalskii2017} models
in nonzero \cite{albarracin2016,krokhmalskii2017} or zero \cite{zhang2016} magnetic field 
were discussed using various complementary approaches.
In particular,
in our recent work \cite{krokhmalskii2017}
it has been shown that the localized-magnon picture \cite{schulenburg2002,zhitomirsky2004,derzhko2015}, 
which emerges for the ideal frustration case,
yields a simple effective description of the low-temperature thermodynamics in a moderate and strong magnetic field.
Since in this regime only the two states on each vertical bond $J_2$,
$\vert u\rangle=\vert\uparrow\uparrow\rangle$
and
$\vert d\rangle=\frac{1}{\sqrt{2}}(\vert\uparrow\downarrow\rangle -\vert\downarrow\uparrow\rangle)$ (localized magnon),
dominate thermodynamic properties,
it is not astonishing that the effective model 
is an antiferromagnetic honeycomb-lattice Ising model in a uniform magnetic field with the Hamiltonian
$H_{\rm{eff}}
=
{\sf{C}}-{\sf{h}}\sum_{m=1}^{\cal{N}}T_m^z
+\sum_{\langle mn\rangle}{\sf{J}}^z T_m^zT_n^z$,
where ${\cal{N}}=N/2$ 
and the (pseudo)spin-$\frac{1}{2}$ operators are defined as follows:
$T^z=\frac{1}{2}(\vert u\rangle\langle u\vert-\vert d\rangle\langle d\vert )$,
$T^+=\vert u\rangle\langle d\vert$,
and
$T^-=\vert d\rangle\langle u\vert$.
Moreover, 
slightly away from the ideal frustration case 
we arrive at an Ising-like $XXZ$ Heisenberg antiferromagnet in an external field along the easy axis on the honeycomb lattice
with the Hamiltonian
\begin{eqnarray}
\label{02}
H_{\rm{eff}}
\!=\!
{\sf{C}}\!-\!{\sf{h}}\!\sum_{m=1}^{\cal{N}}\!T_m^z
\!+\!\sum_{\langle mn\rangle}\left[{\sf{J}}^z T_m^zT_n^z\!+\!{\sf{J}}\left(T_m^xT_n^x+T_m^yT_n^y\right)\right],
\nonumber\\
{\sf{C}}={\cal{N}}\left(-\frac{h}{2}-\frac{J_2}{4}+\frac{3J}{8}\right),
\;\;\;
J=\frac{J_1+J_{\rm{X}}}{2},
\nonumber\\
{\sf{h}}=h-J_2-\frac{3J}{2},
\;\;\;
{\sf{J}}^z=J,
\;\;\;
{\sf{J}}=J_1-J_{\rm{X}}.
\end{eqnarray}
The effective model (\ref{02}) was used in Ref.~\cite{krokhmalskii2017} to explain a peculiarity of the ground-state magnetization curve 
that is related to a spin-flop transition 
which is present in a two-dimensional Ising-like $XXZ$ Heisenberg antiferromagnet in an external field along the easy axis.
However, the magnetothermodynamics of the frustrated honeycomb-lattice bilayer quantum antiferromagnet
(see Eq.~(\ref{01}) and Fig.~\ref{fig1}),
which can be examined on the basis of the effective model (\ref{02}),
was beyond the scope of that paper.
Now we fill this gap
and present results for some low-temperature thermodynamic quantities 
of the frustrated honeycomb-lattice bilayer quantum antiferromagnet in a magnetic field.
It is important to note that, since the quantum spin model (\ref{01}) is frustrated, 
a direct application of quantum Monte Carlo approach is impossible because of the
infamous ``sign problem''.
However,
this powerful method can be applied to the (unfrustrated) effective model (\ref{02}) describing the low-energy degrees of freedom.  

\section{Magnetothermodynamics. Exact diagonalization and quantum Monte Carlo}

We begin with testing the accuracy of the effective-model description (\ref{02}).
To this end,
we consider the full initial model (\ref{01}) on a finite bilayer lattice of $N=24$ sites (see Fig.~3 in Ref.~\cite{krokhmalskii2017})
and perform exact-diagonalization calculations \cite{spinpack} to obtain thermodynamic quantities.  
Then we compare these findings with the results of the exact-diagonalization study of the corresponding effective model (\ref{02}) of ${\cal{N}}=12$ sites.
For concreteness, we fix the set of parameters as follows:
$J_2=5$ 
and 
$J_1=1.1$, $J_{\rm{X}}=0.9$.

First we consider the magnetization curve $M(T,h)$ at zero temperature, see Fig.~\ref{fig1}, bottom.
In case of ideal frustration, 
i.e., $J_1=J_{\rm{X}}=1$, 
the $M(h)/{\cal N}$ curve is independent of the system size:  
$M$ is zero until $h<h_2=J_2=5$, 
it acquires one-half of the saturation value if $h_2<h<h_{\rm{sat}}=J_2+3J_1=8$,
and
achieves the saturation for $h>h_{\rm{sat}}$ \cite{krokhmalskii2017}.
Deviations from the ideal frustration case lead to modifications around $h_2$ and  $h_{\rm{sat}}$, 
however, the wide plateaus are still present, see Fig.~\ref{fig1}, bottom. 

\begin{figure}
\begin{center}
\includegraphics[width=0.45\textwidth]{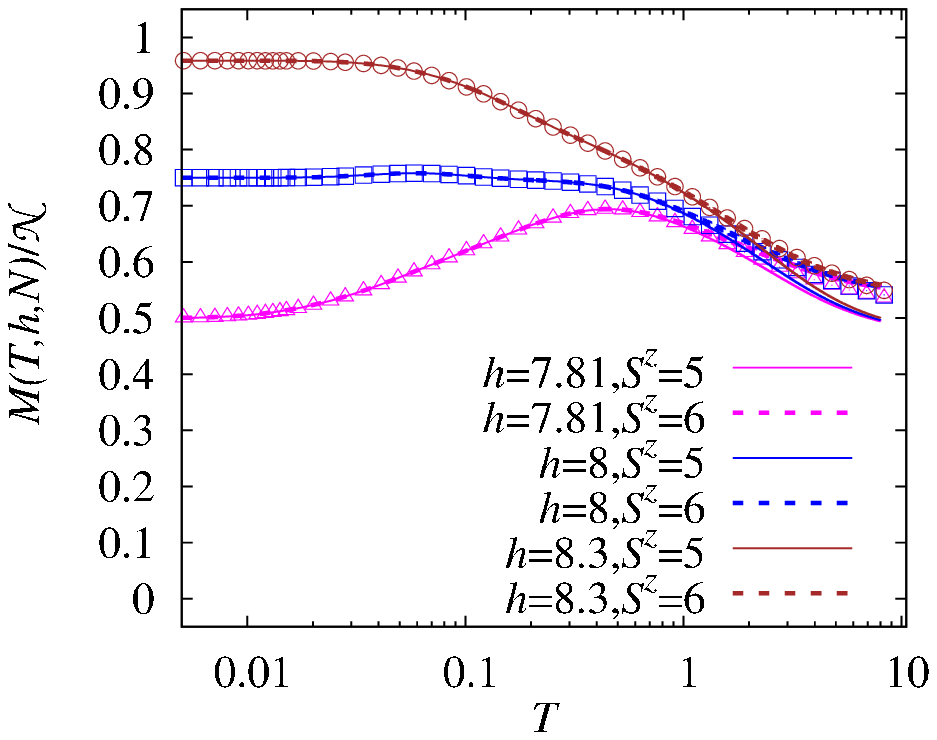}\\
\vspace{5mm}
\includegraphics[width=0.45\textwidth]{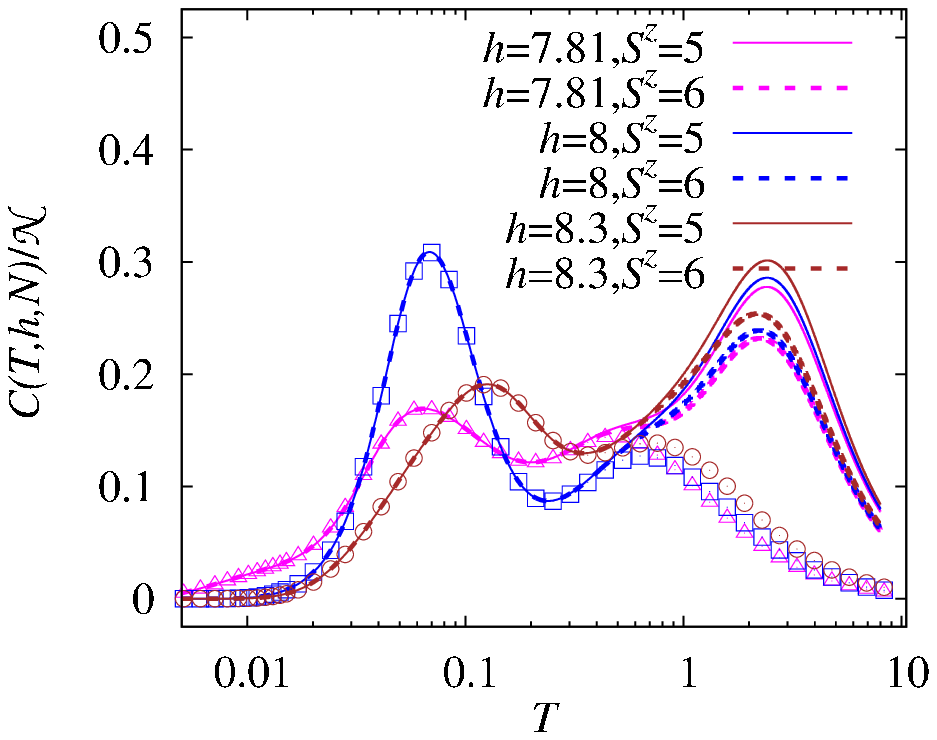}
\end{center}
\vspace{-0.5cm}
\caption{Magnetization (top) and specific heat (bottom) for the system at hand with $J_2=5$, $J_1=1.1$, and $J_{\rm{X}}=0.9$ 
at different fields $h=7.81$ (magenta), $h=8$ (blue), and $h=8.3$ (brown).
Exact-diagonalization results for initial model (\ref{01}) of $N=24$ sites (lines) are compared 
to exact-diagonalization results for effective model (\ref{02}) of ${\cal{N}}=12$ sites (symbols).}
\label{fig2}
\end{figure}

Next we report the temperature dependences of the magnetization and the specific heat for magnetic fields around the saturation field, 
see Fig.~\ref{fig2}.
It is in order to comment the applied exact-diagonalization approach.
The total size of the Hamiltonian matrix for model (\ref{01}) increases as $S^z$ decreases to zero
and becomes beyond the present computational possibilities for $S^z<5$
(even exploiting symmetries already for $S^z=5$ we face a matrix of total size $57\,687\times 57\,687$).
Fortunately,
for the system at hand near the saturation field, 
the subspaces with small $S^z$ becomes relevant at high temperatures only.
This is evident from the comparison of the results in Fig.~\ref{fig2} 
which account the subspaces with $S^z=12,\ldots,5$ (solid lines) and the subspaces with $S^z=12,\ldots,6$ (broken lines).
Clearly,
the exact-diagonalization data for the initial model (\ref{01}) in Fig.~\ref{fig2} are reliable at least up to $T=0.7$.
Furthermore, 
effective-model predictions (symbols) reproduce perfectly all low-temperature features in Fig.~\ref{fig2} for temperatures until about $T=0.5$.
Clearly, 
the temperature region in which effective theory is accurate depends on the values of $J_2$, $J_1$, $J_{\rm{X}}$, and $h$,
however, 
the statement that the simpler (unfrustrated) effective model correctly describes low-energy degrees of freedom is not questioned.
Concerning the temperature profiles of $M(T)$ and $C(T)$ shown in Fig.~\ref{fig2}, 
a prominent feature is the increase of the magnetization as the temperature is growing
as it is found for magnetic fields slightly below $h_{\rm{sat}}$. 
That is related to a large manifold of low-lying states having larger values of total $S^z$ than the ground state, 
and, these states becomes accessible as $T$ increases.    
Another unconventional feature is the double-peak structure of the specific heat. 
Again, a large manifold of low-lying states is responsible, 
however, the value of the total $S^z$ of these states is irrelevant for $C(T)$.

\begin{figure}
\begin{center}
\includegraphics[width=0.425\textwidth]{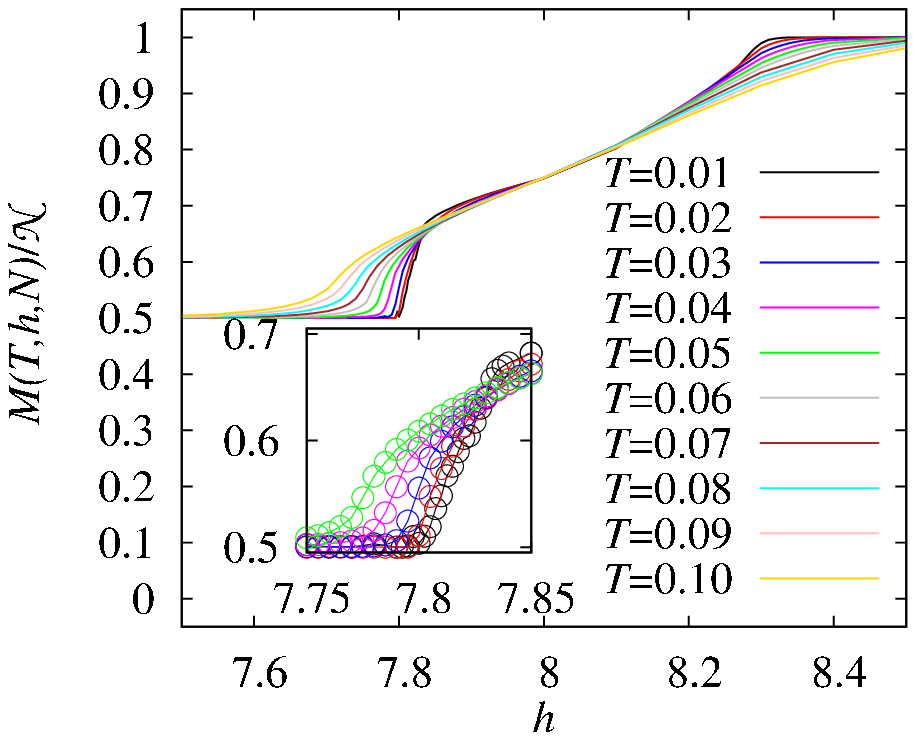}\\
\vspace{5mm}
\includegraphics[width=0.45\textwidth]{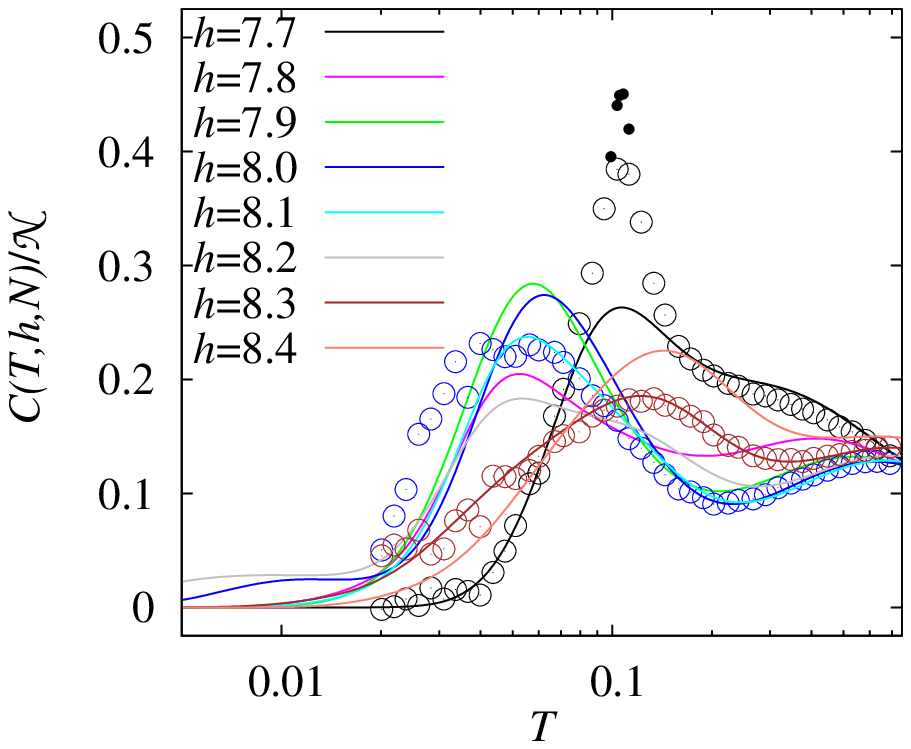}
\end{center}
\vspace{-0.5cm}
\caption{Magnetization curves at different temperatures (top)
and
temperature dependences of the specific heat at different fields (bottom)
for $J_2=5$, $J_1=1.1$, and $J_{\rm{X}}=0.9$
as they follow from 
quantum Monte Carlo simulations 
(${\cal{N}}$ up to $2\,304$, top panel 
and 
${\cal{N}}$ up to $1\,024$, circles in the bottom panel) 
and 
exact diagonalizations (${\cal{N}}=18$, solid lines in the bottom panel) 
for effective model (\ref{02}). 
All curves $C(T,h,N)/{\cal{N}}\to 0$ with $T\to 0$ as it should be.}
\label{fig3}
\end{figure}

Having  shown that the effective model works well at least up to $T = 0.5$,
we  perform 
quantum Monte Carlo calculations \cite{alps} 
and 
exact diagonalizations \cite{spinpack} 
for the (unfrustrated) effective model (\ref{02}) considering much larger systems, 
see Fig.~\ref{fig3}.
The main peculiarity of the magnetization curve shown in Fig.~\ref{fig3} is related to a spin-flopping process present in model (\ref{02}):
antiferromagnetically interacting (pseudo)spins abruptly change their direction 
from parallel to perpendicular orientation with the respect to the easy axis of the anisotropic $XXZ$ model (\ref{02})
at some critical magnetic field $h_{c}$, 
where $h_{c}$ is slightly above 7.8 for the considered set of parameters.
In particular for quantum spins,  
this process discussed for the first time by Louis N\'{e}el in 1936,
is not trivial at all 
depending on the lattice, spin value, temperature fluctuations etc.
We are not aware of studies of the spin-flop phenomenon in the quantum Ising-like $XXZ$ Heisenberg model on a honeycomb lattice
(see, however, some related studies in Ref.~\cite{kohno1997,yunoki2002,holtschneider2007,balamurugan2014})
and such a study goes beyond the scope of this short article.
However, a number of features are obvious from the results reported in Fig.~\ref{fig3}.
Thus,
at sufficiently low temperatures (say, below $T=0.02$) the magnetization around $h_c$ is hardly modified.
But as temperature increases further, 
the value at which magnetization starts to grow rapidly becomes smaller and the slope of magnetization curve becomes smaller too.
Finally, the magnetization becomes moderately rounded 
and at sufficiently high temperature (say, above $T=0.1$) no traces of the spin-flop transition are visible.
The temperature dependence of the specific heat exhibits a low-temperature maximum,
see Figs.~\ref{fig2} and \ref{fig3}.
Within the spin-flop phase, i.e., between $h_c$ and $h_{\rm{sat}}$, excitations are gapless,
but they are gapped outside this field region.
As a result,
the curves $C(T,h,N)$ against $T$ 
exhibit similar low-temperature behavior for $h$ between $h_c$ and $h_{\rm{sat}}$
and differ from such curves for $h$ outside this field region,
see the low-temperature region above $T=0.02$.
Furthermore,
it is interesting to compare $C(T)$ profiles for $h=7.7$ and $h=8.0$ in Fig.~\ref{fig3}.
While the former one reflects a transition from the antiferromagnetic to paramagnetic phase,
the latter one reflects a transition from the spin-flop to paramagnetic phase \cite{holtschneider2007}.
Noticeable finite-size effects for $h=7.7$ 
(large empty circles correspond to ${\cal{N}}=256$ 
whereas small filed circles correspond to ${\cal{N}}=1\,024$)
indicate a singularity which emerges in the thermodynamic limit \cite{krokhmalskii2017}.
In contrast, the temperature-driven transition between the spin-flop and paramagnetic phase 
is not accompanied by a specific-heat singularity.
These traces of the spin-flop phase are expected to be seen for the initial model in the considered parameter region.

\section{Conclusions}

In this paper, 
we have demonstrated that the effective model (\ref{02}) 
can be used to describe the low-temperature thermodynamics 
of the frustrated honeycomb-lattice bilayer quantum antiferromagnet (\ref{01})
around the ideal frustration regime
when $J_2>3(J_1+J_{\rm{X}})/2$ and $\vert J_1-J_{\rm{X}}\vert/J_2\ll 1$.
If deviations from the ideal frustration regime  are present (i.e., for $ J_1
\ne J_{\rm{X}}$),
the magnetization jump transforms into a spin-flop transition
and the model exhibits interesting low-temperature properties
related to the arisen spin-flop phase.
Remarkably, 
the spin-flop physics emerges in the {\it{spin-$\frac{1}{2}$ isotropic}} (i.e., $XXX$) Heisenberg antiferromagnet (\ref{01}),
without any explicit anisotropy,
only due to the lattice geometry and the specific values of exchange couplings $J_{ij}$.

Concerning experimental realizations of the frustrated honeycomb-lattice bilayer spin system, 
the magnetic compound  Bi$_3$Mn$_4$O$_{12}$(NO$_3$) is a candidate,
although, the exchange parameters of the spin Hamiltonian for Bi$_3$Mn$_4$O$_{12}$(NO$_3$) are still under debate \cite{kandpal2011,alaei2017},
and it might happen the $J_2$ does not have sufficient strength.
The search for other honeycomb materials, where our findings would be observable, is desirable and is encouraged.

\end{document}